\documentstyle[12pt]{article}

\begin{document}

\begin{flushright}
DPUR/TH/71\\
April, 2021\\
\end{flushright}
\vspace{20pt}

\pagestyle{empty}
\baselineskip15pt

\begin{center}
{\large\bf  Restricted Weyl Symmetry and Spontaneous Symmetry Breakdown of Conformal Symmetry
\vskip 1mm }

\vspace{20mm}

Ichiro Oda\footnote{
           E-mail address:\ ioda@sci.u-ryukyu.ac.jp
                  }

\vspace{10mm}
           Department of Physics, Faculty of Science, University of the 
           Ryukyus,\\
           Nishihara, Okinawa 903-0213, Japan\\

\end{center}


\vspace{10mm}
\begin{abstract}

We elucidate the relation between the restricted Weyl symmetry and spontaneous symmetry breakdown of
conformal symmetry. Using a scalar-tensor gravity, we show that the restricted Weyl symmetry leads to 
spontaneous symmetry breakdown of a global scale symmetry when the vacuum expectation value of a scalar 
field takes a non-zero value. It is then shown that this spontaneous symmetry breakdown induces spontaneous 
symmetry breakdown of special conformal symmetry in a flat Minkowski space-time, but the resultant 
Nambu-Goldstone boson is not an independent physical mode but expressed in terms of the derivative 
of the dilaton which is the Nambu-Goldstone boson of the global scale symmetry. In other words, the theories 
which are invariant under the general coordinate transformation and the restricted Weyl transformation 
exhibit a Nambu-Goldstone phase where both special conformal transformation and dilatation are spontaneously 
broken while preserving the Poincar\'{e} symmetry.

\end{abstract}

\newpage
\pagestyle{plain}
\pagenumbering{arabic}


\section{Introduction}

It is nowadays widely believed among particle physicists that global and local scale symmetries have conformal (or trace) 
anomaly at the quantum level \cite{Duff}. However, it has been also known since a long time ago that when a scalar field called 
the ``dilaton'' \footnote{Precisely speaking, the dilaton is a Nambu-Goldstone boson of the broken scale invariance \cite{Fujii}.
Here we loosely use this name.} is present and takes a non-zero vacuum expectation value in a theory, there is an ingenious way 
of perturbatively quantizing the theory which preserves the local scale (or Weyl) invariance \cite{Englert}. Thus far, this has been 
several times rediscovered in Refs. \cite{Fradkin}-\cite{Ghilencea2}. The price we have to pay here is that the theory becomes 
non-renormalizable, but this issue is not so serious at least in the presence of the Einstein-Hilbert term of general relativity 
since the Einstein-Hilbert term is non-renormalizable in itself.
 
With the scale invariant regularization scheme \cite{Englert}-\cite{Ghilencea2}, scale symmetries are not explicitly violated 
by the conformal anomaly but it is expected that they might be spontaneously broken as in gauge symmetries in quantum 
field theories, and as a result we could make use of scale symmetries even at the quantum level. In this article, we would
like to discuss spontaneous symmetry breakdown of conformal symmetry in a flat Minkowski space-time. 

In our previous work \cite{Oda-R}, we have already clarified a spontaneous symmetry breakdown of a global scale
symmetry on the basis of a general scalar-tensor gravity with a complex scalar field, a gauge field and higher-derivative
terms which is invariant under the restricted Weyl transformation \cite{Edery1}-\cite{Edery3}. The main purpose of this article 
is to generalize the case of the global scale symmetry to that of special conformal symmetry. We will see later that 
the special conformal symmetry and the global scale symmetry are spontaneously broken while preserving the Poincar\'{e}
symmetry when the dilaton takes a non-zero vacuum expectation value. 

The outline of this paper is as follows: In Section 2, we show that with the nonvanishing dilaton the global scale invariance is 
necessarily spontaneously broken by using the simplest scalar-tensor gravity which is invariant 
under the restricted Weyl transformation in addition to the global scale transformation and the general coordinate transformation. 
In Section 3, we discuss that the restricted Weyl transformation generates conformal transformation in a flat Minkowski 
space-time. This is a nontrivial generalization of the Zumino's theorem in that the full Weyl symmetry is replaced with 
the restricted Weyl symmetry. In Section 4, in the conformal symmetry obtained in Section 3, the special conformal 
symmetry as well as the global scale symmetry are spontaneously broken while the Poincar\'{e} symmetry is kept.

\section{A scalar-tensor gravity}

In this section, we consider the simplest scalar-tensor gravity \cite{Fujii} whose Lagrangian is given by\footnote{We 
follow the conventions and notation of the MTW textbook \cite{MTW}.}
\begin{eqnarray}
{\cal{L}} = \sqrt{-g} \left( \frac{1}{2} \xi \phi^2 R - \frac{1}{2} g^{\mu\nu} \partial_\mu \phi \partial_\nu \phi \right),
\label{ST-gravity}  
\end{eqnarray}
where $\xi$ is a constant called the non-minimal coupling constant, $\phi$ a real scalar field with a normal kinetic term 
(i.e., not a ghost), and $R$ the scalar curvature.
In addition to the general coordinate transformation and a global scale transformation with $\Omega = \textrm{constant}$, 
this Lagrangian is invariant under the following restricted Weyl transformation \cite{Edery1}-\cite{Kamimura}, \cite{Oda-R} 
\begin{eqnarray}
g_{\mu\nu} \rightarrow g^\prime_{\mu\nu} = \Omega^2 (x) g_{\mu\nu}, \qquad 
\Phi \rightarrow \Phi^\prime = \Omega^{-1}(x) \Phi, 
\label{Res-Weyl}  
\end{eqnarray}
where the gauge transformation parameter $\Omega(x)$, which we will call a scale factor, obeys a constraint 
$\Box_g \Omega = 0$.\footnote{To distinguish the difference of the d'Alembertian operators, we will henceforth use 
$\Box_\eta$ for the Minkowski space-time and $\Box_g$ for the Riemannian space-time, respectively.} In order to prove 
the invariance, we need to use the following transformation of the scalar curvature under (\ref{Res-Weyl}):
\begin{eqnarray}
R \rightarrow R^\prime = \Omega^{-2} ( R - 6 \Omega^{-1} \Box_g \Omega ). 
\label{Weyl-R}  
\end{eqnarray}

The field equations obtained from the Lagrangian (\ref{ST-gravity}) read
\begin{eqnarray}
\xi \phi^2 G_{\mu\nu} &=& T_{\mu\nu} - \xi ( g_{\mu\nu} \Box_g - \nabla_\mu \nabla_\nu ) (\phi^2),
\nonumber\\
\xi \phi R + \Box_g \phi &=& 0,
\label{Field-eq}  
\end{eqnarray}
where the Einstein tensor $G_{\mu\nu}$ and the energy-momentum tensor for the scalar field $\phi$ are defined by
\begin{eqnarray}
G_{\mu\nu} &=&  R_{\mu\nu} - \frac{1}{2} g_{\mu\nu} R,
\nonumber\\
T_{\mu\nu} &=& \partial_\mu \phi \partial_\nu \phi - \frac{1}{2} g_{\mu\nu} (\partial_\rho \phi)^2.
\label{G&T}  
\end{eqnarray}
Taking the trace of the former equation in (\ref{Field-eq}) yields
\begin{eqnarray}
\xi \phi^2 R =  (\partial_\mu \phi)^2 + 3 \xi \Box_g (\phi)^2.
\label{Trace-eq}  
\end{eqnarray}
Multiplying the latter equation in (\ref{Field-eq}) by $\phi$ leads to
\begin{eqnarray}
\xi \phi^2 R + \phi \Box_g \phi = 0.
\label{phi-eq}  
\end{eqnarray}
By eliminating $R$ from Eqs. (\ref{Trace-eq}) and (\ref{phi-eq}), we obtain
\begin{eqnarray}
( 6 \xi + 1 ) \Box_g (\phi^2) = 0.
\label{phi-eq2}  
\end{eqnarray}
In this article, we confine ourselves to the case $\xi \neq - \frac{1}{6}$ since in the specific case $\xi = - \frac{1}{6}$
we have a local scale symmetry (or equivalently, the Weyl symmetry). 

Now, in order to understand the physical implications of the constraint $\Box_g \Omega = 0$ of the restricted Weyl transformation,
we are interested in zero-mode solutions to the constraint. It is then obvious that the only zero-mode solution is given by
$\Omega(x) = \textrm{const.}$ which corresponds to a global scale invariance. Thus, let us pay our attention to the scale 
invariance. In an infinitesimal form of the scale factor $\Omega = e^\Lambda$ with $|\Lambda| \ll 1$, the infinitesimal 
gauge transformation parameter $\Lambda$ must obey a constraint $\Box_g \Lambda = 0$ as well, so the zero-mode 
solution is given by 
\begin{eqnarray}
\Lambda (x) = c,
\label{Lambda}  
\end{eqnarray}
where $c$ is a constant. 

Since there is a global invariance associated with the parameter $c$, we can construct a conserved 
Noether charge $Q$. Following the calculation \cite{Fujii, Oda-H}, it turns out that the conserved current for 
the global scale invariance reads
\begin{eqnarray}
J^\mu = \frac{6 \xi + 1}{2} \sqrt{-g} g^{\mu\nu} \partial_\nu (\phi^2).
\label{Noether current}  
\end{eqnarray} 
We can easily verify that this current is conserved, $\partial_\mu J^\mu = 0$, by using the field 
equation (\ref{phi-eq2}). One might wonder why no derivatives of the metric appear in the expression of $J^\mu$.
This is because the derivatives of $\phi$ are mixed with the metric, thus making $J^0$
serve as a generator of the metric transformation. Moreover, in case of a conformal coupling $\xi 
= - \frac{1}{6}$, the conserved current is identically vanishing \cite{Jackiw, Oda-U}. 

Using the corresponding Noether charge defined as $Q = \int d^3 x J^0$, we find that 
\begin{eqnarray}
\delta g_{\mu\nu}  = [ i c Q, g_{\mu\nu} ] = 2 c g_{\mu\nu}, \qquad
\delta \phi  = [ i c Q, \phi ] = - c \phi,
\label{Q-transf1}  
\end{eqnarray}
from which we have
\begin{eqnarray}
[ i Q, g_{\mu\nu} ] = 2 g_{\mu\nu}, \qquad
[ i Q, \phi ] = - \phi.
\label{Q-CR1}  
\end{eqnarray}
Assuming $\langle 0| g_{\mu\nu} |0 \rangle = \eta_{\mu\nu}$\footnote{In particular,  in the presence of the cosmological 
constant, we would need to take a more general fixed background $\bar g_{\mu\nu}$ which satisfies 
$\langle 0| g_{\mu\nu} |0 \rangle = \bar g_{\mu\nu}$.} and $\langle 0| \phi |0 \rangle = \phi_0$ ($\phi_0$ is a constant), 
and then taking the vacuum expectation value of Eq. (\ref{Q-CR1}) leads to
\begin{eqnarray}
\langle 0| [ i Q, g_{\mu\nu} ] |0 \rangle = 2 \eta_{\mu\nu}, \qquad
\langle 0| [ i Q, \phi ] |0 \rangle = - \phi_0.
\label{Q-SSB}  
\end{eqnarray}
Eq. (\ref{Q-SSB}) implies that the global scale invariance must be broken spontaneously for $\phi_0 \neq 0$ at the quantum 
level \cite{Oda-R}. 

Here two remarks are in order. One remark is that in the scale invariant regularization method \cite{Englert}-\cite{Ghilencea2},
by definition the ``dilaton'' cannot vanish anywhere to make all coupling constants be dimensionless. In the typical models 
with spontaneous symmetry breakdown, such a configuration is usually picked up from a potential which induces 
the symmetry breaking, but the existence of the scale invariance in our theory does not allow us to have such a nontrivial potential 
at the classical level. At the quantum level, however, we have a nontrivial effective potential inducing the spontaneous symmetry 
breakdown \cite{Ghilencea1, Ghilencea2}, which is harmony with our assumption $\langle 0| \phi |0 \rangle = \phi_0 \neq 0$ in hand.
As another remark, since the flat Minkowski metric $g_{\mu\nu} = \eta_{\mu\nu}$ is invariant under a combination of 
the general coordinate transformation and the restricted Weyl transformation as will be shown in the next section, 
the former equation in Eq. (\ref{Q-SSB}) is irrelevant to the spontaneous symmetry breakdown of the global scale invariance.  

Next, let us verify explicitly that the spontaneous symmetry breakdown of the global symmetry occurs by moving 
from the Jordan frame (J-frame) to the Einstein frame (E-frame) \cite{Fujii, Oda-P}. To do so, we will move to the Einstein frame 
by implementing a local scale transformation only for the metric tensor except for the scalar field 
\begin{eqnarray}
g_{\mu\nu} \rightarrow g_{\ast\mu\nu} = \Omega^2 (x) g_{\mu\nu}.
\label{Transf-E-frame}  
\end{eqnarray}
Under this scale transformation we have \cite{Fujii}
\begin{eqnarray}
g^{\mu\nu} &=& \Omega^2 (x) g_\ast^{\mu\nu}, \qquad
\sqrt{-g} = \Omega^{-4} \sqrt{-g_\ast}, 
\nonumber\\
R &=& \Omega^2 ( R_\ast + 6 \Box_\ast f - 6 g_\ast^{\mu\nu} f_\mu f_\nu ), 
\label{Transf-E-frame2}  
\end{eqnarray}
where we have defined
\begin{eqnarray}
f \equiv \log \Omega, \quad
\Box_\ast f \equiv \frac{1}{\sqrt{- g_\ast}} \partial_\mu ( \sqrt{- g_\ast} g_\ast^{\mu\nu} \partial_\nu f), \quad
f_\mu \equiv \partial_\mu f = \frac{\partial_\mu \Omega}{\Omega}.
\label{f}  
\end{eqnarray}

Then, the Lagrangian density (\ref{ST-gravity}) is cast to the form
\begin{eqnarray}
{\cal{L}} = \sqrt{-g_\ast} \Biggl[ \frac{1}{2} \xi \phi^2 \Omega^{-2} ( R_\ast + 6 \Box_\ast f 
- 6 g_\ast^{\mu\nu} f_\mu f_\nu ) - \frac{1}{2} \Omega^{-2} g_\ast^{\mu\nu} \partial_\mu \phi \partial_\nu \phi \Biggr].
\label{E-Lag}  
\end{eqnarray}
To reach the Einstein frame, we have to choose the scale factor $\Omega(x)$ to satisfy the relation
\begin{eqnarray}
\Omega^2 = \frac{1}{M_{Pl}^2} \xi \phi^2,
\label{Planck-mass}  
\end{eqnarray}
where $M_{Pl}$ is the reduced Planck mass. Note that Eq. (\ref{Planck-mass}) shows that the ``dilaton'' $\phi$ cannot vanish 
in this case either since we cannot move to the E-frame from the J-frame in the case of the vanishing dilaton. As a result, we obtain 
a Lagrangian in the E-frame:
\begin{eqnarray}
{\cal{L}} &=& \sqrt{-g_\ast} \Biggl( \frac{M_{Pl}^2}{2} R_\ast - \frac{1}{2} g_\ast^{\mu\nu} 
\partial_\mu \sigma \partial_\nu \sigma \Biggr).
\label{E-Lag2}  
\end{eqnarray}
Here we have defined a scalar field $\sigma(x)$ and a constant $\zeta$ as
\begin{eqnarray}
\phi = \xi^{- \frac{1}{2}} M_{Pl} \, e^{\frac{\zeta}{M_{Pl}} \sigma}, \qquad
\zeta = \sqrt{\frac{\xi}{6 \xi + 1}}. 
\label{phi-sigma}  
\end{eqnarray}

The Lagrangian (\ref{E-Lag2}) indicates that the scalar field $\sigma(x)$ is a massless Nambu-Goldstone field
associated with the spontaneous symmetry breakdown of the global scale invariance. Indeed, the current $J^\mu$
in Eq.  (\ref{Noether current}) can be rewritten in terms of $\sigma(x)$ as
\begin{eqnarray}
J^\mu = \frac{M_{Pl}}{\zeta} \sqrt{-g_\ast} g_\ast^{\mu\nu} \partial_\nu \sigma.
\label{Noether current2}  
\end{eqnarray} 
The corresponding charge is given by
\begin{eqnarray}
Q = \int d^3 x J^0 = \frac{M_{Pl}}{\zeta}  \int d^3 x \sqrt{-g_\ast} g_\ast^{0\nu} \partial_\nu \sigma.
\label{Noether charge}  
\end{eqnarray} 
Since $Q$ has a linear term in $\sigma(x)$, it is obvious that the charge cannot annihilate the vacuum $| 0 \rangle$
\begin{eqnarray}
Q | 0 \rangle \neq 0,
\label{charge-vac}  
\end{eqnarray} 
which implies the spontaneous symmetry breakdown of the scale symmetry. 

This fact can be also verified by evaluating the vacuum expectation value of the commutator between $Q$ and 
$\sigma(x)$. Actually, from the Lagrangian (\ref{E-Lag2}) the canonical conjugate momentum for the scalar field 
$\sigma(x)$ reads
\begin{eqnarray}
\pi_\sigma \equiv \frac{\partial {\cal{L}}}{\partial \partial_0 \sigma} = - \sqrt{-g_\ast} g_\ast^{0\nu} \partial_\nu \sigma.
\label{CCM}  
\end{eqnarray} 
Then, the Noether charge $Q$ in Eq. (\ref{Noether charge}) can be rewritten as 
\begin{eqnarray}
Q = - \frac{M_{Pl}}{\zeta}  \int d^3 x \, \pi_\sigma.
\label{Noether charge-pi}  
\end{eqnarray} 
Using the equal-time commutation relation 
\begin{eqnarray}
[ \sigma (t, \vec{x}), \pi_\sigma  (t, \vec{y}) ] = i \delta^3 (x - y),
\label{ET-CR}  
\end{eqnarray} 
we obtain
\begin{eqnarray}
[ i Q, \sigma(x) ] = - \frac{M_{Pl}}{\zeta}.
\label{Q-sigma}  
\end{eqnarray} 
Taking the vacuum expectation value of this equation yields
\begin{eqnarray}
\langle 0 | [ i Q, \sigma(x) ] | 0 \rangle = - \frac{M_{Pl}}{\zeta} \neq 0,
\label{Noether charge-VEV}  
\end{eqnarray} 
which clearly means that the scalar field $\sigma(x)$ is the Nambu-Goldstone boson for the scale symmetry. 
Note that because of the definition (\ref{phi-sigma}), the spontaneous symmetry breakdown of the scale symmetry 
in the J-frame can be interpreted as that of the shift symmetry in the E-frame.

\section{Conformal symmetry from the restricted Weyl symmetry}

Let us recall that conformal transformation \cite{Gross} can be defined as the general coordinate transformation which can be undone 
by the Weyl transformation when the space-time metric is the flat Minkowski metric. In the scalar-tensor gravity (\ref{ST-gravity}) 
there is no Weyl invariance, but instead we have the restricted Weyl invariance, so we could define the conformal transformation
by replacing the Weyl transformation with the restricted Weyl one. With this definition, the conformal transformation is described 
by the equation 
\begin{eqnarray}
\partial_\mu \epsilon_\nu + \partial_\nu \epsilon_\mu = 2 \Lambda(x) \eta_{\mu\nu},
\label{Conf-Killing}  
\end{eqnarray} 
where the infinitesimal scale factor $\Lambda(x)$ obeys the constraint $\Box_\eta \Lambda = 0$.   

Taking the trace of Eq. (\ref{Conf-Killing}) enables us to determine $\Lambda(x)$ to be
\begin{eqnarray}
\Lambda = \frac{1}{4} \partial^\rho \epsilon_\rho.
\label{lambda}  
\end{eqnarray} 
Inserting this $\Lambda$ to Eq. (\ref{Conf-Killing}) yields 
\begin{eqnarray}
\partial_\mu \epsilon_\nu + \partial_\nu \epsilon_\mu = \frac{1}{2} \partial^\rho \epsilon_\rho \eta_{\mu\nu},
\label{Conf-Killing2}  
\end{eqnarray} 
which is often called the ``conformal Killing equation'' in the Minkowski space-time. It is worth stressing that
Eq. (\ref{Conf-Killing2}) implies the following fact: The flat Minkowski metric $g_{\mu\nu} = \eta_{\mu\nu}$
is invariant in the space of the metric functions under a suitable combination of the general coordinate transformation 
and the restricted Weyl transformation such that  
\begin{eqnarray}
\delta ( \epsilon_\mu ) = \delta_{GCT} ( \epsilon_\mu ) -  \delta_{RW} ( \Lambda = \frac{1}{4} \partial^\rho \epsilon_\rho ),
\label{Comb-transf}  
\end{eqnarray} 
when the vector field $\epsilon_\mu(x)$ obeys the conformal Killing equation (\ref{Conf-Killing2}). To put it differently, 
the characteristic feature of the theory under consideration is that the Lagrangian (\ref{ST-gravity}) possesses 
the conformal symmetry with 15 global parameters which is a subgroup of the general coordinate transformation 
and the restricted Weyl transformation.

Multiplying it by $\partial^\mu \partial^\nu$, we obtain
\begin{eqnarray}
\Box_\eta \partial^\rho \epsilon_\rho = 0.
\label{Conf-Killing3}  
\end{eqnarray} 
Moreover, multiplying Eq. (\ref{Conf-Killing2}) by $\partial^\mu \partial_\lambda$ and then symmetrizing the 
indices $\nu$ and $\lambda$ leads to the desired equation
\begin{eqnarray}
\partial_\lambda \partial_\nu \partial^\rho \epsilon_\rho = 0,
\label{Conf-Killing4}  
\end{eqnarray} 
where we have used Eqs. (\ref{Conf-Killing2}) and (\ref{Conf-Killing3}). It turns out that a general solution to 
Eq. (\ref{Conf-Killing4}) reads
\begin{eqnarray}
\epsilon^\mu = a^\mu + \omega^{\mu\nu} x_\nu + \lambda x^\mu + b^\mu x^2 - 2 x^\mu b_\rho x^\rho,
\label{Conf-Killing-vector}  
\end{eqnarray} 
where $a^\mu, \omega^{\mu\nu} = - \omega^{\nu\mu}, \lambda$ and $b^\mu$ are all constant parameters and 
they correspond to the translation, the Lorentz transformation, the dilatation\footnote{For clarity, we will call 
a global scale transformation in a flat Minlowski space-time ``dilatation''. Dilatation is usually interpreted as 
a subgroup of the general coordinate transformation in a such way that the space-time coordinates are transformed 
as $x^\mu \rightarrow \Omega x^\mu$ in the flat space-time where $\Omega$ is a constant scale factor, 
whereas the global scale transformation is a rescaling of all lengths by the same $\Omega$ by 
$g_{\mu\nu} \rightarrow \Omega^2 g_{\mu\nu}$. The two viewpoints are completely equivalent 
since all the lengths are defined via the line element $d s^2 = g_{\mu\nu} d x^\mu d x^\nu$.} and 
the special conformal transformation, respectively.
 
At this stage, we have to verify that the infinitesimal scale factor $\Lambda$ generated by the ``conformal Killing vector'' 
$\epsilon^\mu$ in Eq. (\ref{Conf-Killing-vector}) satisfies the constraint of the restricted Weyl symmetry.
Actually, substituting Eq. (\ref{Conf-Killing-vector}) into Eq. (\ref{lambda}), we have
\begin{eqnarray}
\Lambda = \lambda - 2 b_\mu x^\mu.
\label{lambda2}  
\end{eqnarray} 
This is nothing but zero-mode solutions to the constraint $\Box_\eta \Lambda = 0$, so the restricted Weyl
transformation can generate the conformal transformation in a flat Minkowski background as in the full Weyl transformation.

To summarize, we have shown that the restricted Weyl symmetry, together with the general coordinate
invariance, generates the conformal symmetry in the flat Minkowski background. This result is a generalization of 
the well-known Zumino's theorem \cite{Zumino} which insists that the theories invariant under both the general coordinate
transformation and the Weyl transformation (or local scale transformation) possess conformal transformation in the
flat Minkowski background. In the theory in hand, the Weyl transformation is replaced with the restricted Weyl transformation
which is a subgroup of the Weyl transformation.

\section{Spontaneous symmetry breakdown of special conformal symmetry}

In this section, we will show that in the conformal transformation obtained in the previous section through the restricted 
Weyl transformation, both the special conformal transformation and dilatation are spontaneously broken.
 
Before doing so, let us recall the generators and algebra of conformal transformation \cite{Gross}. The generators of the translation
($P_\mu$), the Lorentz transformation ($M_{\mu\nu}$), the dilatation ($D$) and the special conformal transformation 
($K_\mu$) respectively take the form
\begin{eqnarray}
P_\mu &=& - i \partial_\mu, \qquad 
M_{\mu\nu} = - i ( x_\mu \partial_\nu - x_\nu \partial_\mu ),
\nonumber\\
D &=& i x^\mu \partial_\mu, \qquad
K_\mu = i ( 2 x_\mu x^\rho \partial_\rho - x^2 \partial_\mu ).
\label{CF-gen}  
\end{eqnarray} 
The conformal group is a simple Lie group and its algebra reads
\begin{eqnarray}
&{}& [ M_{\mu\nu}, M_{\rho\sigma} ] = i ( \eta_{\mu\rho} M_{\nu\sigma} -  \eta_{\nu\rho} M_{\mu\sigma} 
+ \eta_{\mu\sigma} M_{\rho\nu} -  \eta_{\nu\sigma} M_{\rho\mu}  ),
\nonumber\\
&{}& [ P_\mu, M_{\rho\sigma} ] = - i ( \eta_{\mu\rho} P_\sigma - \eta_{\mu\sigma} P_\rho ), \qquad
[ P_\mu, P_\nu ] = 0,  \qquad
[ P_\mu, D ] = i P_\mu,  
\nonumber\\
&{}& [ M_{\mu\nu}, D ] = 0, \qquad
[ K_\mu, D ] = - i K_\mu, \qquad
[ P_\mu, K_\nu ] = - 2i ( \eta_{\mu\nu} D + M_{\mu\nu} ), 
\nonumber\\
&{}& [ K_\mu, K_\nu ] = 0, \qquad
[ K_\mu, M_{\rho\sigma} ] = - i ( \eta_{\mu\rho} K_\sigma - \eta_{\mu\sigma} K_\rho ).   
\label{CF-algebra}  
\end{eqnarray} 
 
We assume that the vacuum is invariant under the Poincar\'{e} transformation
\begin{eqnarray}
P_\mu | 0 \rangle = 0, \qquad 
M_{\mu\nu} |0 \rangle = 0.
\label{Poin-vac}  
\end{eqnarray} 
In Section 2, we have already proved that the global scale transformation is spontaneously broken. In particular, 
setting $g_{\mu\nu} = \eta_{\mu\nu}$, the global scale transformation, i.e., the dilatation, must be spontaneously broken 
as well.  The corresponding Nambu-Goldstone boson is the massless dilaton $\sigma(x)$ as seen 
in Eq. (\ref{Noether charge-VEV}), which gives us the equation
\begin{eqnarray}
\langle 0 | [ D, \sigma(x) ] | 0 \rangle \equiv v \neq 0,
\label{D-sigma}  
\end{eqnarray} 
where $v$ is a non-zero constant.

Now let us consider the Jacobi identity \cite{Kobayashi}
\begin{eqnarray}
\Bigl[ [ P_\mu, K_\nu ], \sigma \Bigr] + \Bigl[ [ K_\nu, \sigma ], P_\mu \Bigr] 
+ \Bigl[ [ \sigma, P_\mu ], K_\nu \Bigr] = 0.
\label{Jacobi}  
\end{eqnarray} 
Next, using the conformal algebra for $[ P_\mu, K_\nu ]$ in Eq. (\ref{CF-algebra}) and the equation
\begin{eqnarray}
[ P_\mu, \sigma ] = - i \partial_\mu \sigma,
\label{Heisen-eq}  
\end{eqnarray} 
we can obtain
\begin{eqnarray}
- 2i [ \eta_{\mu\nu} D + M_{\mu\nu}, \sigma ] + \Bigl[ [ K_\nu, \sigma ], P_\mu \Bigr] + i [ \partial_\mu \sigma, K_\nu ] = 0.
\label{Jacobi2}  
\end{eqnarray} 
Taking the vacuum expectation value of the algebra (\ref{Jacobi2}) and using Eqs. (\ref{Poin-vac}) and (\ref{D-sigma})
provides us with the final result
\begin{eqnarray}
\langle 0 | [ K_\nu, \partial_\mu \sigma ] | 0 \rangle =  - 2 v \eta_{\mu\nu}.
\label{Jacobi3}  
\end{eqnarray} 

Eq. (\ref{Jacobi3}) clearly shows that the special conformal transformation is also spontaneously broken, and the corresponding
Nambu-Goldstone boson can be identified with $\partial_\mu \sigma$ \cite{Kobayashi}. Thus, the Nambu-Goldstone boson 
associated with the spontaneous symmetry breakdown of the special conformal transformation is not an independent mode 
but expressed in terms of the derivative of the Nambu-Goldstone boson of the dilatation or the global scale 
transformation \cite{Ivanov}.

\section{Conclusion}

In this paper, based on a scalar-tensor gravity, we have elucidated the relation between the restricted Weyl symmetry 
and spontaneous symmetry breakdown of conformal symmetry. What we have shown is that the theories which are 
simultaneously invariant under the general coordinate transformation and the restricted Weyl transformation have conformal 
transformation when we take the flat Minkowski metric $g_{\mu\nu} = \eta_{\mu\nu}$, but both special conformal transformation 
and dilatation are spontaneously broken. This fact should be contrasted to the Zumino's theorem which insists that the theories 
which are invariant under the general coordinate transformation and the Weyl transformation necessarily have 
conformal transformation without any spontaneous symmetry breakdown when we take the flat Minkowski metric. 

An interesting point is that the Nambu-Goldstone boson of the special conformal transformation is not an independent field 
but the derivative of the Nambu-Goldstone boson of the dilatation. This phenomenon has been already observed in Refs. 
\cite{Ivanov, Kobayashi}, but for the first time we have presented a concrete gravitational model which realizes this
phenomenon by starting with the restricted Weyl transformation. Although we have derived the above results in terms of 
a specific theory, i.e., the scalar-tensor gravity, we believe that the same phenomenon could occur even in more general 
and realistic theories if they have both the general coordinate invariance and the restricted Weyl symmetry. In this respect, 
it is worthwhile to point out that the restricted Weyl symmetry plays an important role since only the global scale symmetry, 
which often coexists with the restricted Weyl symmetry, cannot generate the conformal group owing to the lack of 
the special conformal transformation.

In series of papers related to the restricted Weyl symmetry \cite{Oda-R, Kamimura}, we have investigated some aspects
of the theories with the restricted Weyl symmetry such as the meaning of the constraint, the origin of the restricted Weyl 
symmetry and the spontaneous symmetry breakdown of global symmetries etc. In these papers, it was mentioned that 
the restricted Weyl symmetry is very similar to the restricted gauge symmetry in QED which emerges as a residual symmetry 
when fixing the gauge symmetry by the Lorenz gauge $\partial^\mu A_\mu = 0$ in the sense that the both symmetries
have a constraint on the transformation parameter (in the restricted Weyl symmetry $\Box_g \Omega =0$ whereas 
in the restricted gauge one $\Box_\eta \theta = 0$), and they play a role in spontaneous symmetry breakdown of 
global symmetries. In future, we wish to construct a more realistic model beyond the standard model on the basis of 
such the restricted symmetries.

\begin{flushleft}
{\bf Acknowledgements}
\end{flushleft}

We would like to thank T. Kugo for valuable discussions and information on Ref. \cite{Kobayashi}.
This work is partly supported by the JSPS Kakenhi Grant No. 21K03539.


\end{document}